\documentclass[a4paper]{raa_twocolumn}            
\usepackage{CJK}
\usepackage{graphicx,times}             
\usepackage{natbib}
\usepackage{amssymb,amsmath}
\bibpunct{(}{)}{;}{a}{}{,}

\usepackage[a4paper=true,pagebackref=true]{hyperref}
\hypersetup{colorlinks = true, linkcolor = blue, anchorcolor = red, citecolor = blue, filecolor = red, pagecolor = red, urlcolor = red}

\usepackage{multirow}
\newcommand{\kms}{{\rm km\,s^{-1}}}
\newcommand{\tdd}{{\rm TD/D}}

\defcitealias{Hamer:2024}{HS24}

\begin{document}
\begin{CJK*}{UTF8}{gbsn}

    \title{Are Near Resonant Multiple-planet Systems from Kepler Young?}
    
   \volnopage{Vol.0 (20xx) No.0, 000--000}      
   \setcounter{page}{1}          

   \author{Wei Zhu (祝伟)\inst{1} \and
   Qingru Hu (胡清茹)\inst{1}}

   \institute{Department of Astronomy, Tsinghua University, Beijing 100084, China; {\it weizhu@tsinghua.edu.cn} \\
\vs\no
   {\small Received~~20xx month day; accepted~~20xx~~month day}}

\abstract
{Recent studies have claimed that Kepler multi-planet systems hosting near-resonant planet pairs---particularly those near second-order mean-motion resonances (MMRs)---exhibit smaller stellar velocity dispersions than the general population of Kepler planet hosts. Interpreting velocity dispersion as an age indicator, these works concluded that near-resonant systems are systematically younger. We revisit this claim, but we explicitly account for contamination by thick disk stars, which are kinematically hotter and follow a different age-velocity dispersion relation (AVR) than thin disk stars. Using the kinematic criterion to separate thin and thick disk stars, we show that systems classified as having plausible second-order resonant pairs are preferentially hosted by brighter, closer stars and are therefore less contaminated by thick disk stars than the overall sample. After applying a cut to remove probable thick disk contaminants ($\tdd<0.1$), the vertical velocity dispersion of near-resonant systems becomes statistically indistinguishable from that of the overall Kepler multi-planet sample. We conclude that the apparent kinematic youth of near-resonant systems in Kepler may not be due to a genuine age difference, but rather arises from observational selection effects linked to host star properties and planet detectability. We also comment on the kinematic ages of ultra-short-period planets (USPs).
}

   \authorrunning{Zhu \& Hu}            
   \titlerunning{Kinematic Age of Near-resonant Systems}  

   \maketitle
%

%
%
\section{Introduction}           
\label{sect:intro}


Mean-motion resonance (MMR) is an appealing feature in the study of multi-planet systems. It occurs when the orbital periods of two planets are close to some small integer ratio and the resonance angle librates. According to the current theoretical understanding, planets can often \citep[e.g.,][]{Borderies:1984, Lee:2002}, though not always \citep[e.g.,][]{Goldreich:2014, Yang:2024, chen:2025}, be locked into MMRs if they undergo convergent migration in the protoplanetary disk, and population synthesis models indeed predict that MMRs should be abundant in multi-planet systems when they are young \citep[e.g.,][]{Izidoro:2017, Emsenhuber:2021}. On the other hand, observations have shown that, among the mature multi-planet systems, those hosting MMR planet pairs are a minor fraction \citep[e.g.,][]{Fabrycky:2014, Huang:2023, Dai:2024}. Growing evidence also appears that even the planet pairs with period ratios very close to small integer ratios are in fact near but not in MMR \citep{Goldberg:2023, Hu:2025}.

Recent observational studies have sought to link the presence of near-resonant architectures to stellar age \citep{Dai:2024, Hamer:2024, Schmidt:2024}. In particular, \citet[hereafter HS24]{Hamer:2024} analyzed a large sample of Kepler multi-planet systems and claimed that systems hosting planet pairs near second-order MMRs and those near first-order MMRs but affected by stellar tides have smaller velocity dispersions than the general field stars hosting planets. Taking the velocity dispersion as an age indicator, those authors argued that systems with near resonant configurations are on average younger. \citet{Schmidt:2024} further converted the measured velocity dispersion into stellar age and claimed that Kepler multi-planet systems with plausible 2nd-order MMRs are only 1--2\,Gyr old, whereas the general Kepler systems span a much wider age range of 1--10\,Gyr \citep[but see][]{Bouma:2024}. The same claim was also made for systems with plausible 1st-order MMRs, but only for systems that are likely affected by tidal dissipation inside the innermost planet \citep{Schmidt:2024}.

While stellar kinematics has been widely accepted as an empirical aging technique, the standard age--velocity dispersion relation \citep[AVR,][]{stromger1946, wielen1977} uses the vertical velocity dispersion and only applies to thin disk stars \citep[e.g.,][]{seabroke2007, aumer2016, Mackereth:2019}, and even so, it appears to vary with Galactic locations \citep[e.g.,][]{ting2019, sharma2021fundamental, sun2025age}. Whereas in \citetalias{Hamer:2024}, the authors used the total velocity dispersion as an age indicator,
\footnote{As pointed out in Hu et al. (in prep), the definition of the velocity standard deviation in \citetalias{Hamer:2024} is non-standard.}
and applied it to the stars in the Kepler field, which span a wide range in distance, Galactic height, and kinematic environment. If a sample is contaminated by thick disk stars, which are kinematically hotter and do not appear to have a well established relation between age and the total velocity dispersion (see \citealt{SilvaAguirre:2018} and \citealt{Miglio:2021} for close examples in the Kepler field), the inferred age of a subsample could be biased. While \citetalias{Hamer:2024} had tried to remove thick disk (and halo) stars from their sample based on the orbital parameters in the Galactic potential, this dynamical method is model-dependent, relies on very precise astrometric and radial velocity measurements, and is not able to identify all (probable) thick disk stars \citep{Alinder:2025}. As an example, Kepler-444 (KIC 6278762) is a known thick disk star based on both its chemical and kinematic properties \citep{Campante:2015}, but it was not identified by the dynamical method of \citetalias{Hamer:2024}.

Besides the dynamical method, other methods are available to separate stars belonging to the different Galactic components, most importantly, the thin and thick disks (see \citealt{Alinder:2025} for an overview). Since most of the stars in the Kepler field do not have $[\alpha/\rm Fe]$ measurements, an appropriate alternative is then the kinematic method, which evaluates the probability of a star belonging to the thick disk relative to the thin disk, often denoted as $\tdd$, based on its 3D velocity \citep[e.g.,][]{Bensby:2003, Bensby:2014}
\begin{equation}
    \tdd \equiv \frac{X_{\rm TD}}{X_{\rm D}} \cdot \frac{f_{\rm TD}}{f_{\rm D}} .
\end{equation}
Here $X_{\rm TD}=0.09$ and $X_{\rm D}=0.85$, which measure the observed fractions of the thick and thin populations in the Solar neighborhood, respectively. The factor $f$ measures the three-dimensional velocity ellipsoid
\begin{equation}
    f = k \cdot \exp\left(
    - \frac{U_{\rm LSR}^2}{2\sigma_U^2}
    - \frac{(V_{\rm LSR}-V_{\rm asym})^2}{2\sigma_V^2}
    - \frac{W_{\rm LSR}^2}{2\sigma_W^2} \right) ,
\end{equation}
where $k\equiv (2\pi)^{-3/2}(\sigma_U \sigma_V \sigma_W)^{-1}$. We follow \citet{Bensby:2014} and adopt characteristic velocity dispersions $(\sigma_U, \sigma_V, \sigma_W)=(35, 20, 16)\,\kms$ and $(67, 38, 35)\,\kms$ for the thin and the thick populations, respectively. The asymmetric drift of the $U$ component has been set at zero, and the asymmetric drift of the $V$ component is chosen as $V_{\rm asym}=-15\,\kms$ for thin disk and $-46\,\kms$ for thick disk.
\footnote{\citet{Chen:2021} proposed an improved $\tdd$ formula that takes into account the density or velocity variations of the thin vs.\ thick disk population across the Galactic distance. We confirm here that the adoption of the \citet{Chen:2021} $\tdd$ formulism does not affect our results.}
Compared to the dynamical method that integrates the stellar orbit for typically a few Gyr, the kinematic method is much easier to implement.

Using this kinematical selection criterion, we show that the planetary host sample of \citetalias{Hamer:2024} is indeed contaminated by (possible) thick disk stars and that the level of contamination varies across the subsets of planetary hosts that were investigated. After a proper correction of the thick disk contamination, we find that the systems with plausible MMR planet pairs are consistent in stellar vertical velocity dispersion with the overall sample of planetary hosts. In other words, our refined analysis does not find statistical evidence supporting the claim of \citetalias{Hamer:2024} that near resonant systems in the Kepler sample are systematically younger than the general population.

This paper is organized in the following way: in Section~\ref{sect:sample} we reconstruct the stellar sample, which follows closely the procedures in \citetalias{Hamer:2024}; in Section~\ref{sec:result} we present our refined analysis and the key results that near-resonant systems do not appear kinematically younger than the general population; in Section~\ref{sec:reason} we present a plausible reason why the near-resonant systems are less contaminated by thick disk stars; in Section~\ref{sec:summary} we provide a brief summary of our findings.

\section{Sample Reconstruction} \label{sect:sample}

\begin{figure*}
    \centering
    \includegraphics[width=0.9\linewidth]{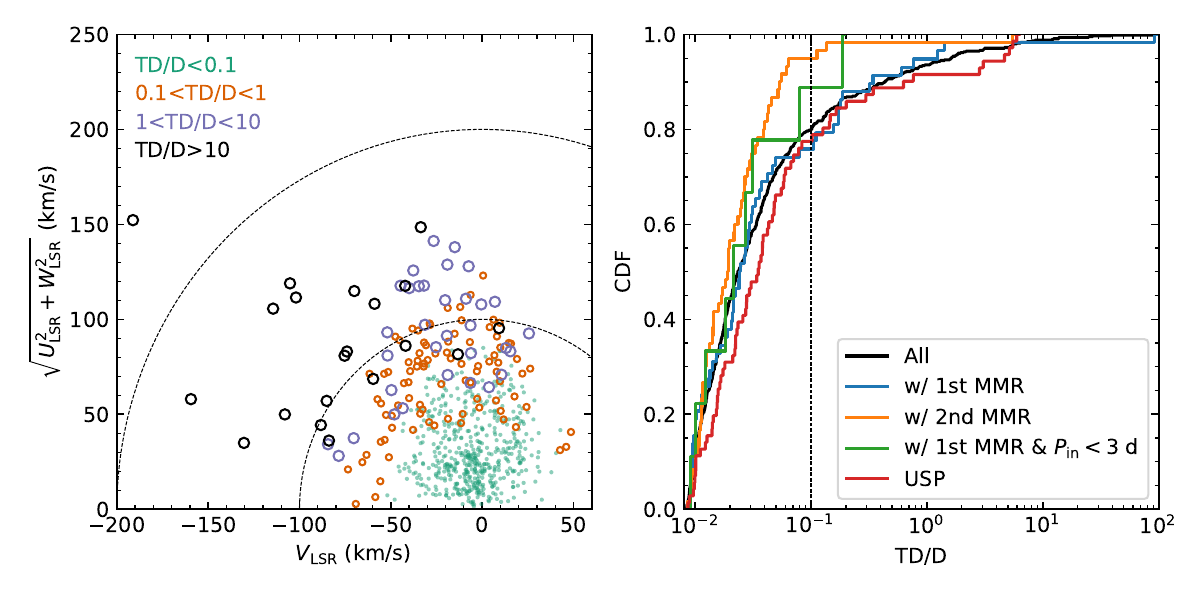}
    \caption{Left panel: Toomre diagram for the stars in our sample, color-coded by the $\tdd$ value. Right panel: cumulative distribution function (CDF) of the $\tdd$ values for stars with planetary systems classified into different categories. A system with at least one planet pair that is identified as plausible 1st-order (2nd-order) resonance by \citetalias{Hamer:2024} is labeled ``w/ 1st MMR'' (``w/ 2nd MMR''). The category, ``w/ 1st MMR \& $P_{\rm in}<3\,$d'', corresponds approximately to the stars with plausibly 1st-order resonant and tidal decay timescale $\tau_e<20$\,Myr in \citetalias{Hamer:2024}. ``USP'' is the sample of planetary systems hosting ultra-short-period planets as identified in \citet{Schmidt:2024}.}
    \label{fig:kinematics}
\end{figure*}

We start from the Table~1 of \citet{Schmidt:2024}, which provides names and the adopted RV values of the systems used in \citetalias{Hamer:2024}. Systems with in/near 1st-order and 2nd-order MMR pairs have been labeled, which were defined using the $\delta_{\rm res}$ criterion of \citetalias{Hamer:2024}. After cross-matching with the Cumulative KOI Data of NASA Exoplanet Archive \citep{Christiansen:2025}, we obtain the Kepler-band magnitudes and the observed transit multiplicity of the systems. In this step, we additionally exclude 16 stars that were reported as USP hosts in \citet{Schmidt:2024}, but which we identify as false positives.
The final sample therefore contains 667 planetary systems, which include 60 labeled with plausible 1st-order MMR pairs, 60 with plausible 2nd-order MMR pairs, and nine that are labeled as tidally affected 1st-order MMR systems. 
Similar to \citetalias{Hamer:2024}, the parallax and proper motion measurements are taken from Gaia DR3 \citep{gaia_dr3}, and the stellar kinematics are computed via \texttt{pyia} \citep{pyia}.

The classification of stars into different Galactic components affects the conversion from stellar kinematics to their ages. We classify the stars in our sample into thin and thick disk components based on the $\tdd$ parameter, which is the relative probability that a star belongs to the thin disk over the thick disk \citep{Bensby:2014}.

The left panel of Figure~\ref{fig:kinematics} shows the Toomre diagram for the stars in our sample. While the majority of the stars have kinematics resembling those of typical thin disk stars, some do act more like thick disk stars, with the derived $\tdd>10$. The cumulative distribution of the $\tdd$ parameter, as shown in the right panel of Figure~\ref{fig:kinematics}, also confirms this. Additionally, while the plausible 1st-order resonant systems do follow the overall stellar distribution and have a long tail towards large $\tdd$, almost all of the plausible 2nd-order resonant systems belong to the thin disk component with very few exceptions (i.e., three with $\tdd>0.1$). A two-sample Kolmogorov--Smirnov (KS) test between the plausible 2nd-order resonant sample and the overall sample yields $p=0.011$, and the Anderson--Darling (AD) test yields $p=0.007$, both rejecting the hypothesis that the two distributions are drawn from the same underlying distribution. In other words, the systems classified by \citetalias{Hamer:2024} as plausibly in/near 2nd-order MMR are much less contaminated by the (possible) thick disk stars than the systems classified to have plausible 1st-order MMRs. This is the reason that makes the 2nd-order MMR systems appear kinematically younger, as further explained later.

The same is true for the systems classified as plausibly in/near 1st-order MMR and affected stellar tides. To identify systems that belong to this category, we adopt a more empirical criterion rather than adopting the existing label in \citet{Schmidt:2024}. In the calculation of the tidal circularization timescale $\tau_e$ in \citetalias{Hamer:2024}, values of some unknown parameters (e.g., tidal quality factor $Q$) have been assumed, which have (large) systematic uncertainties.
Making use of the strong dependence of the tidal circularization timescale on the orbital period \citepalias[e.g., Figure~3 of][]{Hamer:2024}, we simply choose systems with the innermost orbital period $P_{\rm in}<3\,$d as those with tidal circularization timescale $\tau_e\lesssim 20\,$Myr. With the assumed Mars-like dissipative properties of \citetalias{Hamer:2024}, these values correspond to a typical Kepler planet with a radius of $\sim1.5\,R_\oplus$ and a mass of $\sim5\,M_\oplus$ around a Sun-like host \citep{Lithwick:2012}.
As shown in the right panel of Figure~\ref{fig:kinematics}, systems with plausible 1st-order resonant pairs and $P_{\rm in}<3\,$d are also preferentially around stars that likely belong to the thin disk population, although because of the much smaller sample size the statistical difference is not as large for the 2nd-order resonant systems. As such, the cause of such systems appearing kinematically younger in \citetalias{Hamer:2024} is the same as those with 2nd-order MMRs. Given the uncertainty in the identification of such systems and the small (i.e., nine) number of systems in this sample, we no longer consider this group separately in the following analysis.

In the right panel of Figure~\ref{fig:kinematics} we also show the cumulative distribution of the $\tdd$ parameters derived for the ultra-short-period planet (USP) hosts in \citet{Schmidt:2024}. Unlike the systems with plausible 2nd-order MMRs, USP hosts are more, though at less significant level, contaminated by probable thick disk stars compared to the overall planetary hosts, as has been noticed in \citet{Tu:2025}. However, we do not see statistically significant difference in velocity dispersion between USP hosts and the overall planetary sample, as further explained in Appendix~\ref{sec:usp}.

\section{Near-resonant Kepler systems do not appear kinematically younger} \label{sec:result}

\begin{figure}
    \centering
    \includegraphics[width=\linewidth]{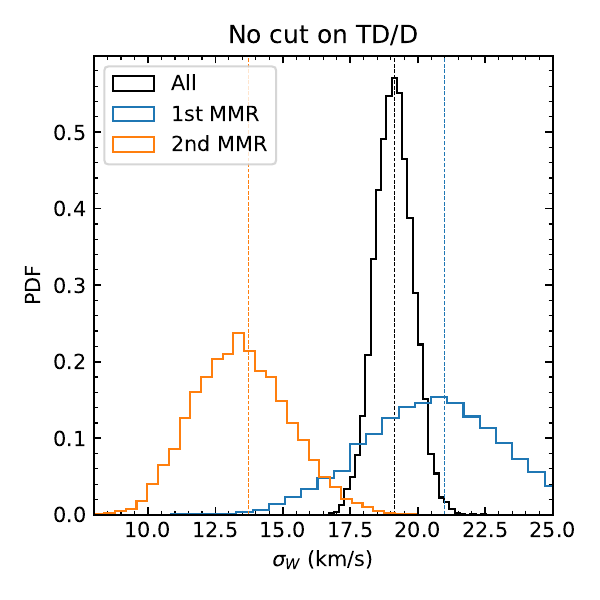}
    \caption{Distributions of the vertical velocity dispersion ($\sigma_W$) from bootstrap tests of the three planet host groups. Here we have included all stars and do not perform cut on $\tdd$. In this case, the velocity dispersion in systems with plausible 2nd-order MMRs is statistically smaller than that of the overall sample, whereas the velocity dispersion in systems with plausible 1st-order MMRs is statistically similar to that of the overall sample. These results are qualitatively similar to that of \citetalias{Hamer:2024}.}
    \label{fig:sigmaw-full}
\end{figure}

\begin{figure*}
    \centering
    \includegraphics[width=0.45\linewidth]{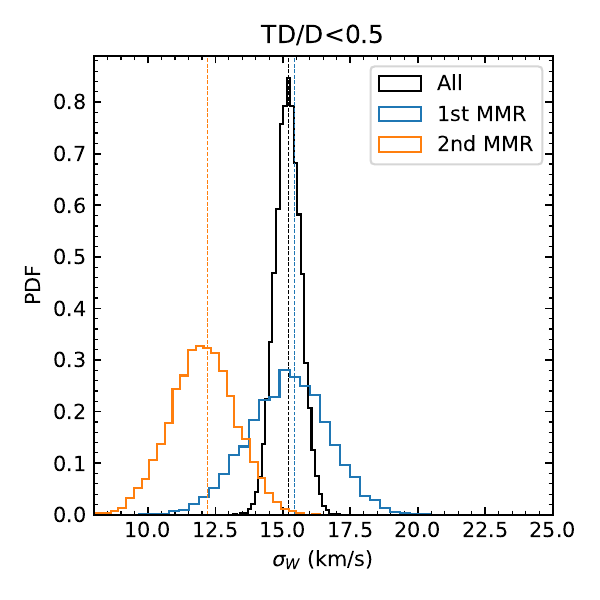}
    \includegraphics[width=0.45\linewidth]{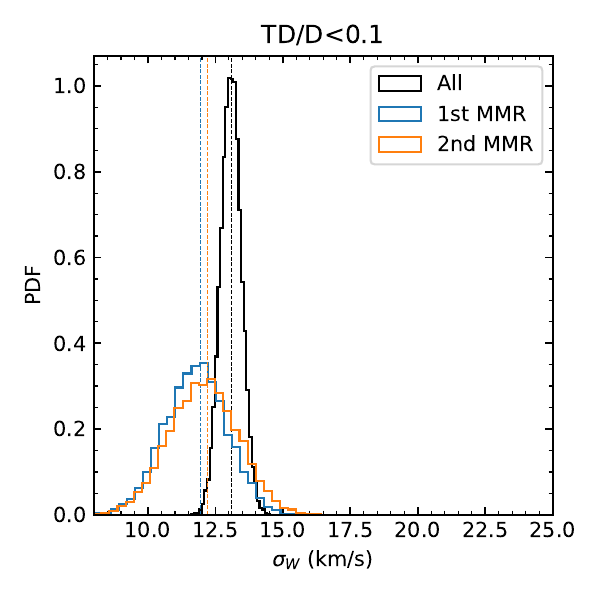}
    \caption{Similar to Figure~\ref{fig:sigmaw-full}, but now we have removed the possible thick disk contaminants from the analysis. In the left panel, stars with $\tdd>0.5$ are excluded, and the statistical difference between the velocity dispersions in stars with plausible 2nd-order MMRs and the overall stellar sample is reduced to $2.1\sigma$. In the right panel, stars with $\tdd>0.1$ are excluded. This stricter cut only removes three out of 60 stars in the plausible 2nd-order MMR group (see right panel of Figure~\ref{fig:kinematics}), but the three resulting velocity dispersions become all statistically similar ($<0.5\sigma$).}
    \label{fig:sigmaw-tdd}
\end{figure*}

We derive the vertical velocity dispersions for the three stellar groups and evaluate their uncertainties via bootstrap tests,
\footnote{These bootstrap tests are done via \texttt{scipy.stats.bootstrap}.}
following closely the procedures of \citetalias{Hamer:2024}. The result before correcting for the thick disk star contaminants is shown in Figure~\ref{fig:sigmaw-full}. Without any cut on the stellar sample, the systems with plausible 2nd-order MMRs do have a smaller velocity dispersion than the overall sample. This reproduces the key result of \citetalias{Hamer:2024}, even though we have slightly different stellar samples and use the more proper age indicator.

Because the Kepler multi-planet systems are contaminated by thick disk stars, which do not follow the same AVR as the thin disk stars, it is reasonable (and compulsory) to remove the contaminants in order to ensure that they do not affect the subsequent statistical analyses. 

We have chosen two different criteria for removing thick disk contaminants, and the results are shown in Figure~\ref{fig:sigmaw-tdd}. In the left panel, we adopt a rather loose criterion of $\tdd<0.5$ as thin disk stars, whereas in the right panel we adopt a stricter criterion of $\tdd<0.1$ as thin disk stars, both having been used in the literature \citep[e.g.,][]{Bensby:2014, Chen:2021, Boettner:2024}. After imposing such cuts on the stellar sample, the velocity dispersion of stars with plausible 2nd-order MMRs is nearly unaffected, but the velocity dispersions of the other two stellar samples are substantially reduced. In particular, only three out of 60 stars in the 2nd-order MMR group is removed by the stricter cut of $\tdd<0.1$, but the resulting velocity dispersion is now statistically indistinguishable to the velocity dispersions of both the overall star sample and the stars with plausible 1st-order MMRs.
\footnote{The smaller standard deviation of the ``All'' sample is simply due to its larger sample size.}
This means that contamination from thick disk stars indeed has impact on the derived velocity dispersions. After such contaminants are removed, the systems with plausible 1st-order or 2nd-order MMRs are both statistically similar in kinematics, and thus in the inferred ages, to the overall sample of stars. 

\citetalias{Hamer:2024} also claimed that the near/in 1st-order MMR systems affected by tidal dissipation inside the innermost planets were kinematically colder than the parent sample. As shown in Figure~\ref{fig:kinematics}, this subset of systems is also less contaminated by the thick disk stars similar to the 2nd-order MMR systems, so their apparently colder kinematics---and thus younger age---can also be explained by the same bias as for the 2nd-order MMR systems.

\begin{figure*}
    \centering
    \includegraphics[width=\linewidth]{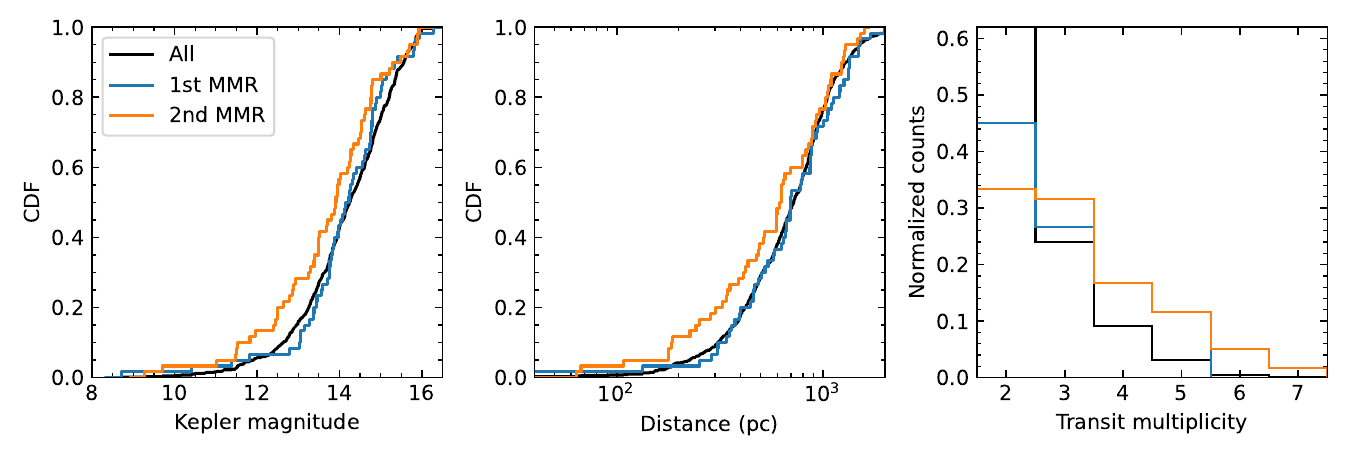}
    \caption{Distributions of the Kepler magnitudes (left panel), distance (middle panel), and observed transit multiplicity (right panel) of the overall stellar sample, the plausible 1st-order MMR systems, and the plausible 2nd-order MMR systems. Compared to the overall stellar sample, stars with plausible 2nd-order MMR planet pairs are systematically brighter, closer, and known to host more transiting planet detections.}
    \label{fig:compare-samples}
\end{figure*}

\section{Why are near-resonant systems less contaminated by thick disk stars?} \label{sec:reason}

We have shown in the previous section that the apparent kinematic youth of near-resonant systems disappears after removing thick disk contaminants, but why are systems with plausible 2nd-order MMRs preferentially found around thin disk stars? The answer lies not in an intrinsic age difference, but (more likely) in observational selection effects linked to the host star properties and the detectability of multiple, small transiting planets.

In Figure~\ref{fig:compare-samples}, we show three stellar/planetary properties across the overall sample, systems with plausible 1st-order MMRs, and systems with plausible 2nd-order MMRs. Stars hosting 2nd-order resonant pairs are systematically brighter and closer than the overall sample (as well as the stars hosting 1st-order resonant pairs). As the photometric precision increases with stellar brightness, one naturally expects to detect smaller and more transiting planets around those brighter stars for the same transit signal-to-noise ratios \citep{Zhu:2020}. This is confirmed by the third panel of Figure~\ref{fig:compare-samples}, which shows that systems with plausible 2nd-order resonant pairs do preferentially contain more transiting planet detections. With more detected planets, the by-chance probability increases that two of the detected planets are closer to some small integer ratio in orbital periods, even if the periods are randomly distributed. Smaller planets can be put into tighter orbits and have smaller period ratios without affecting the long-term stability \citep[e.g., Section~2.4.2 of][]{Zhu:2021}, and the chance also increases that such period ratios hit one of the small integer ratios, especially those corresponding to 2nd-order MMRs. A comprehensive validation of this hypothesis would require the construction of synthetic multi-planet system populations whose statistical properties reproduce those of the observed ensemble. Such an investigation lies beyond the scope of the present study.

As thick disk stars are known to be systematically fainter and more distant than thin disk stars in the Kepler field, the preference for brighter and closer stars in the detection of 2nd-order resonant planet pairs can lead to a sample of stars that are less contaminated by thick disk stars.

\section{Summary} \label{sec:summary}

Using the multi-planet systems from Kepler, a previous work by \citetalias{Hamer:2024} claimed that systems with planets in/near MMRs were kinematically colder and thus younger than the overall multi-planet systems \citep[see also][]{Schmidt:2024}. We show here that the apparent kinematic youth of near-resonant (especially near 2nd-order resonant) systems from the Kepler sample can be explained by the fact that their host stars are preferentially drawn from the brighter and closer thin disk population. After excluding the thick disk contaminants from the sample, which follow a different age--velocity dispersion relation, the vertical velocity dispersions---and hence the inferred ages---of near-resonant systems become statistically indistinguishable from those of the overall Kepler multi-planet sample. Although the occurrence rate of near-resonant planetary systems within the overall exoplanet population may decrease with stellar age \citep{Dai:2024}, our results suggest that this rate ceases to undergo significant evolution beyond a stellar age of approximately $\sim 1$\,Gyr.

Together with Hu et al. (submitted), who re-examined the kinematic ages of hot Jupiter systems, our work highlights the need for caution when using stellar kinematics to infer (relative) ages of planetary systems at the population level.

\begin{acknowledgements}
We thank the anonymous reviewer for comments on the manuscript.
Work by W.Z.\ and Q.H.\ was funded by the National Natural Science Foundation of China (NSFC) under No.\ 12133005.
\end{acknowledgements}

\appendix
\section{On the kinematic ages of USP hosts} \label{sec:usp}

\begin{figure*}
    \includegraphics[width=\textwidth]{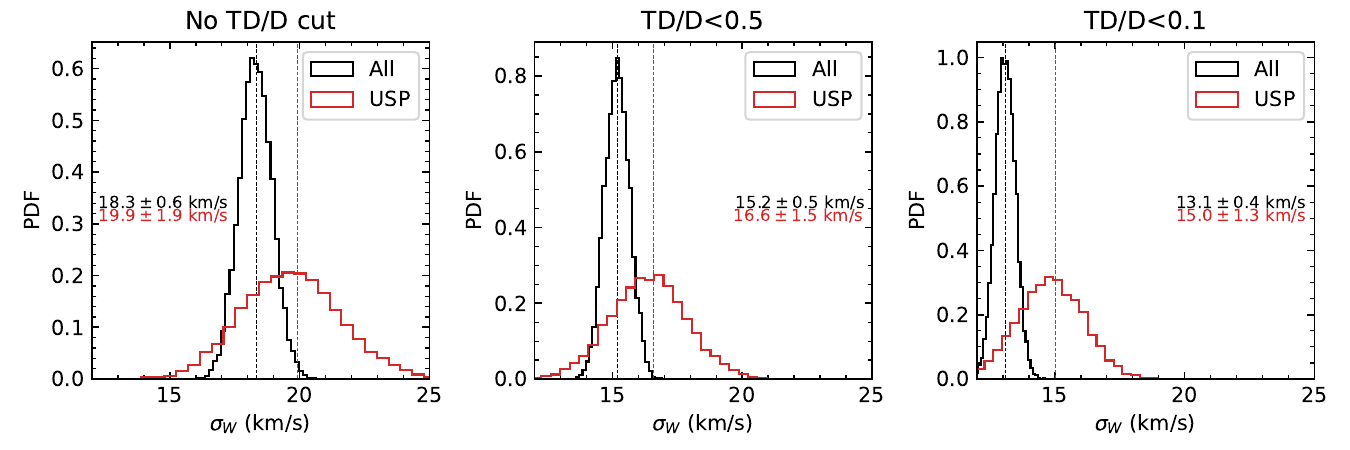}
    \caption{Similar to Figures~\ref{fig:sigmaw-full} and \ref{fig:sigmaw-tdd}, but for the USP hosts in the sample of \citet{Schmidt:2024}. The vertical velocity dispersions and the associated uncertainties estimated from bootstrap tests are shown in each panel. A marginally significant ($\sim2\sigma$) $\sigma_W$ difference is only seen when all stars with $\tdd>0.1$ are excluded. If no $\tdd$ cut or a less strict cut ($\tdd<0.5$) is imposed, the $\sigma_W$ values derived from the USP hosts and the overall planetary hosts are statistically indistinguishable ($<1\sigma$).} \label{fig:usp}
\end{figure*}

We briefly comment on the kinematic ages of USP hosts here. As shown in Figure~\ref{fig:kinematics} and noted in Section~\ref{sect:sample}, a larger fraction of USP hosts belong to the (probable) thick disk, although the statistical difference is marginal at $p=0.06$ for both KS and AD tests. However, as shown in Figure~\ref{fig:usp}, the derived vertical velocity dispersion $\sigma_W$ shows no significant difference between the USP hosts and the overall planetary hosts. It is only when a strict cut of $\tdd<0.1$ is imposed that we see a marginal $\sigma_W$ difference at the level of $\sim 2\sigma$. Here we have not corrected for the other differences in the USP host properties \citep[e.g., stellar types;][]{Tu:2025}. 

As the result is sensitive to the selection criterion of thin disk stars, it is unclear whether the ages of the USP hosts are indeed systematically older than the general planetary hosts. We leave a more systematic analysis of the USP systems to some future study.

\bibliographystyle{raa}
\bibliography{bibtex}

@ARTICLE{SilvaAguirre:2018,
       author = {{Silva Aguirre}, V. and {Bojsen-Hansen}, M. and {Slumstrup}, D. and {Casagrande}, L. and {Kawata}, D. and {Ciuc{\v{a}}}, I. and {Handberg}, R. and {Lund}, M.~N. and {Mosumgaard}, J.~R. and {Huber}, D. and {Johnson}, J.~A. and {Pinsonneault}, M.~H. and {Serenelli}, A.~M. and {Stello}, D. and {Tayar}, J. and {Bird}, J.~C. and {Cassisi}, S. and {Hon}, M. and {Martig}, M. and {Nissen}, P.~E. and {Rix}, H.~W. and {Sch{\"o}nrich}, R. and {Sahlholdt}, C. and {Trick}, W.~H. and {Yu}, J.},
        title = "{Confirming chemical clocks: asteroseismic age dissection of the Milky Way disc(s)}",
      journal = {\mnras},
         year = 2018,
        month = apr,
       volume = {475},
       number = {4},
        pages = {5487-5500},
          doi = {10.1093/mnras/sty150},
archivePrefix = {arXiv},
       eprint = {1710.09847},
 primaryClass = {astro-ph.GA},
       adsurl = {https://ui.adsabs.harvard.edu/abs/2018MNRAS.475.5487S}
}

@ARTICLE{Dai:2024,
       author = {{Dai}, Fei and {Goldberg}, Max and {Batygin}, Konstantin and {van Saders}, Jennifer and {Chiang}, Eugene and {Choksi}, Nick and {Li}, Rixin and {Petigura}, Erik A. and {Gilbert}, Gregory J. and {Millholland}, Sarah C. and {Dai}, Yuan-Zhe and {Bouma}, Luke and {Weiss}, Lauren M. and {Winn}, Joshua N.},
        title = "{The Prevalence of Resonance Among Young, Close-in Planets}",
      journal = {\aj},
         year = 2024,
        month = dec,
       volume = {168},
       number = {6},
          eid = {239},
        pages = {239},
          doi = {10.3847/1538-3881/ad83a6},
archivePrefix = {arXiv},
       eprint = {2406.06885},
 primaryClass = {astro-ph.EP},
       adsurl = {https://ui.adsabs.harvard.edu/abs/2024AJ....168..239D}
}

@ARTICLE{Hamer:2024,
       author = {{Hamer}, Jacob H. and {Schlaufman}, Kevin C.},
        title = "{Kepler-discovered Multiple-planet Systems near Period Ratios Suggestive of Mean-motion Resonances Are Young}",
      journal = {\aj},
         year = 2024,
        month = feb,
       volume = {167},
       number = {2},
          eid = {55},
        pages = {55},
          doi = {10.3847/1538-3881/ad110e},
archivePrefix = {arXiv},
       eprint = {2312.02260},
 primaryClass = {astro-ph.EP},
       adsurl = {https://ui.adsabs.harvard.edu/abs/2024AJ....167...55H}
}

@software{pyia,
    author       = {Adrian Price-Whelan and
                    Gabe Brammer},
    title        = {pyia},
    month        = jul,
    year         = 2021,
    publisher    = {Zenodo},
    version      = {v1.3},
    doi          = {10.5281/zenodo.5057363},
    url          = {https://doi.org/10.5281/zenodo.5057363}
}

@ARTICLE{Bensby:2014,
       author = {{Bensby}, T. and {Feltzing}, S. and {Oey}, M.~S.},
        title = "{Exploring the Milky Way stellar disk. A detailed elemental abundance study of 714 F and G dwarf stars in the solar neighbourhood}",
      journal = {\aap},
         year = 2014,
        month = feb,
       volume = {562},
          eid = {A71},
        pages = {A71},
          doi = {10.1051/0004-6361/201322631},
archivePrefix = {arXiv},
       eprint = {1309.2631},
 primaryClass = {astro-ph.GA},
       adsurl = {https://ui.adsabs.harvard.edu/abs/2014A&A...562A..71B}
}

@ARTICLE{Christiansen:2025,
       author = {{Christiansen}, Jessie L. and {McElroy}, Douglas L. and {Harbut}, Marcy and {Ciardi}, David R. and {Crane}, Megan and {Good}, John and {Hardegree-Ullman}, Kevin K. and {Kesseli}, Aurora Y. and {Lund}, Michael B. and {Lynn}, Meca and {Muthiar}, Ananda and {Nilsson}, Ricky and {Oluyide}, Toba and {Papin}, Michael and {Rivera}, Amalia and {Swain}, Melanie and {Susemiehl}, Nicholas D. and {Tam}, Raymond and {van Eyken}, Julian and {Beichman}, Charles},
        title = "{The NASA Exoplanet Archive and Exoplanet Follow-up Observing Program: Data, Tools, and Usage}",
      journal = {\psj},
         year = 2025,
        month = aug,
       volume = {6},
       number = {8},
          eid = {186},
        pages = {186},
          doi = {10.3847/PSJ/ade3c2},
archivePrefix = {arXiv},
       eprint = {2506.03299},
 primaryClass = {astro-ph.EP},
       adsurl = {https://ui.adsabs.harvard.edu/abs/2025PSJ.....6..186C}
}

@ARTICLE{Chen:2021,
       author = {{Chen}, Di-Chang and {Xie}, Ji-Wei and {Zhou}, Ji-Lin and {Dong}, Subo and {Liu}, Chao and {Wang}, Hai-Feng and {Xiang}, Mao-Sheng and {Huang}, Yang and {Luo}, Ali and {Zheng}, Zheng},
        title = "{Planets Across Space and Time (PAST). I. Characterizing the Memberships of Galactic Components and Stellar Ages: Revisiting the Kinematic Methods and Applying to Planet Host Stars}",
      journal = {\apj},
         year = 2021,
        month = mar,
       volume = {909},
       number = {2},
          eid = {115},
        pages = {115},
          doi = {10.3847/1538-4357/abd5be},
archivePrefix = {arXiv},
       eprint = {2102.09424},
 primaryClass = {astro-ph.EP},
       adsurl = {https://ui.adsabs.harvard.edu/abs/2021ApJ...909..115C}
}

@ARTICLE{Boettner:2024,
       author = {{Boettner}, C. and {Viswanathan}, A. and {Dayal}, P.},
        title = "{Exoplanets across galactic stellar populations with PLATO: Estimating exoplanet yields around FGK stars for the thin disk, thick disk, and stellar halo}",
      journal = {\aap},
         year = 2024,
        month = dec,
       volume = {692},
          eid = {A150},
        pages = {A150},
          doi = {10.1051/0004-6361/202451537},
archivePrefix = {arXiv},
       eprint = {2407.15917},
 primaryClass = {astro-ph.EP},
       adsurl = {https://ui.adsabs.harvard.edu/abs/2024A&A...692A.150B}
}

@ARTICLE{Zhu:2020,
       author = {{Zhu}, Wei},
        title = "{On the Patterns Observed in Kepler Multi-planet Systems}",
      journal = {\aj},
         year = 2020,
        month = may,
       volume = {159},
       number = {5},
          eid = {188},
        pages = {188},
          doi = {10.3847/1538-3881/ab7814},
archivePrefix = {arXiv},
       eprint = {1907.02074},
 primaryClass = {astro-ph.EP},
       adsurl = {https://ui.adsabs.harvard.edu/abs/2020AJ....159..188Z}
}

@ARTICLE{Zhu:2021,
       author = {{Zhu}, Wei and {Dong}, Subo},
        title = "{Exoplanet Statistics and Theoretical Implications}",
      journal = {\araa},
         year = 2021,
        month = sep,
       volume = {59},
        pages = {291-336},
          doi = {10.1146/annurev-astro-112420-020055},
archivePrefix = {arXiv},
       eprint = {2103.02127},
 primaryClass = {astro-ph.EP},
       adsurl = {https://ui.adsabs.harvard.edu/abs/2021ARA&A..59..291Z}
}

@ARTICLE{Schmidt:2024,
       author = {{Schmidt}, Stephen P. and {Schlaufman}, Kevin C. and {Hamer}, Jacob H.},
        title = "{Resonant and Ultra-short-period Planet Systems Are at Opposite Ends of the Exoplanet Age Distribution}",
      journal = {\aj},
         year = 2024,
        month = sep,
       volume = {168},
       number = {3},
          eid = {109},
        pages = {109},
          doi = {10.3847/1538-3881/ad5d76},
archivePrefix = {arXiv},
       eprint = {2407.04765},
 primaryClass = {astro-ph.EP},
       adsurl = {https://ui.adsabs.harvard.edu/abs/2024AJ....168..109S}
}

@ARTICLE{Bouma:2024,
       author = {{Bouma}, Luke G. and {Hillenbrand}, Lynne A. and {Howard}, Andrew W. and {Isaacson}, Howard and {Masuda}, Kento and {Palumbo}, Elsa K.},
        title = "{Ages of Stars and Planets in the Kepler Field Younger than Four Billion Years}",
      journal = {\apj},
         year = 2024,
        month = dec,
       volume = {976},
       number = {2},
          eid = {234},
        pages = {234},
          doi = {10.3847/1538-4357/ad855f},
archivePrefix = {arXiv},
       eprint = {2410.06246},
 primaryClass = {astro-ph.SR},
       adsurl = {https://ui.adsabs.harvard.edu/abs/2024ApJ...976..234B}
}

@ARTICLE{wielen1977,
       author = {{Wielen}, R.},
        title = "{The Diffusion of Stellar Orbits Derived from the Observed Age-Dependence of the Velocity Dispersion}",
      journal = {\aap},
         year = 1977,
        month = sep,
       volume = {60},
       number = {2},
        pages = {263-275},
       adsurl = {https://ui.adsabs.harvard.edu/abs/1977A&A....60..263W}
}

@ARTICLE{stromger1946,
       author = {{Str{\"o}mberg}, Gustaf},
        title = "{The Motions of the Stars Within 20 Parsecs of the Sun.}",
      journal = {\apj},
         year = 1946,
        month = jul,
       volume = {104},
        pages = {12},
          doi = {10.1086/144830},
       adsurl = {https://ui.adsabs.harvard.edu/abs/1946ApJ...104...12S}
}

@article{seabroke2007,
  title={Revisiting the relations: Galactic thin disc age--velocity dispersion relation},
  author={Seabroke, GM and Gilmore, G},
  journal={Monthly Notices of the Royal Astronomical Society},
  volume={380},
  number={4},
  pages={1348--1368},
  year={2007},
  publisher={The Royal Astronomical Society}
}

@ARTICLE{ting2019,
       author = {{Ting}, Yuan-Sen and {Rix}, Hans-Walter},
        title = "{The Vertical Motion History of Disk Stars throughout the Galaxy}",
      journal = {\apj},
         year = 2019,
        month = jun,
       volume = {878},
       number = {1},
          eid = {21},
        pages = {21},
          doi = {10.3847/1538-4357/ab1ea5},
archivePrefix = {arXiv},
       eprint = {1808.03278},
 primaryClass = {astro-ph.GA},
       adsurl = {https://ui.adsabs.harvard.edu/abs/2019ApJ...878...21T}
}

@ARTICLE{aumer2016,
       author = {{Aumer}, Michael and {Binney}, James and {Sch{\"o}nrich}, Ralph},
        title = "{Age-velocity dispersion relations and heating histories in disc galaxies}",
      journal = {\mnras},
         year = 2016,
        month = oct,
       volume = {462},
       number = {2},
        pages = {1697-1713},
          doi = {10.1093/mnras/stw1639},
archivePrefix = {arXiv},
       eprint = {1607.01972},
 primaryClass = {astro-ph.GA},
       adsurl = {https://ui.adsabs.harvard.edu/abs/2016MNRAS.462.1697A}
}

@article{sharma2021fundamental,
  title={Fundamental relations for the velocity dispersion of stars in the Milky Way},
  author={Sharma, Sanjib and Hayden, Michael R and Bland-Hawthorn, Joss and Stello, Dennis and Buder, Sven and Zinn, Joel C and Kallinger, Thomas and Asplund, Martin and De Silva, Gayandhi M and D’Orazi, Valentina and others},
  journal={Monthly Notices of the Royal Astronomical Society},
  volume={506},
  number={2},
  pages={1761--1776},
  year={2021},
  publisher={Oxford University Press}
}

@article{sun2025age,
  title={The Age--Velocity Dispersion Relations of the Galactic Disk as Revealed by the LAMOST-Gaia Red Clump Stars},
  author={Sun, Weixiang and Shen, Han and Jiang, Biwei and Liu, Xiaowei},
  journal={The Astrophysical Journal},
  volume={979},
  number={2},
  pages={103},
  year={2025},
  publisher={The American Astronomical Society}
}

@ARTICLE{Alinder:2025,
       author = {{Alinder}, Simon and {Bensby}, Thomas and {McMillan}, Paul},
        title = "{Impact of selection criteria on the structural parameters of the Galactic thin and thick discs}",
      journal = {arXiv e-prints},
         year = 2025,
        month = nov,
          eid = {arXiv:2511.10092},
        pages = {arXiv:2511.10092},
          doi = {10.48550/arXiv.2511.10092},
archivePrefix = {arXiv},
       eprint = {2511.10092},
 primaryClass = {astro-ph.GA},
       adsurl = {https://ui.adsabs.harvard.edu/abs/2025arXiv251110092A}
}

@ARTICLE{Miglio:2021,
       author = {{Miglio}, A. and {Chiappini}, C. and {Mackereth}, J.~T. and {Davies}, G.~R. and {Brogaard}, K. and {Casagrande}, L. and {Chaplin}, W.~J. and {Girardi}, L. and {Kawata}, D. and {Khan}, S. and et al.},
        title = "{Age dissection of the Milky Way discs: Red giants in the Kepler field}",
      journal = {\aap},
         year = 2021,
        month = jan,
       volume = {645},
          eid = {A85},
        pages = {A85},
          doi = {10.1051/0004-6361/202038307},
archivePrefix = {arXiv},
       eprint = {2004.14806},
 primaryClass = {astro-ph.GA},
       adsurl = {https://ui.adsabs.harvard.edu/abs/2021A&A...645A..85M}
}

@ARTICLE{Bensby:2003,
       author = {{Bensby}, T. and {Feltzing}, S. and {Lundstr{\"o}m}, I.},
        title = "{Elemental abundance trends in the Galactic thin and thick disks as traced by nearby F and G dwarf stars}",
      journal = {\aap},
         year = 2003,
        month = nov,
       volume = {410},
        pages = {527-551},
          doi = {10.1051/0004-6361:20031213},
       adsurl = {https://ui.adsabs.harvard.edu/abs/2003A&A...410..527B}
}

@ARTICLE{Borderies:1984,
       author = {{Borderies}, N. and {Goldreich}, P.},
        title = "{A Simple Derivation of Capture Probabilities for the J+1:J and J+2:J Orbit-Orbit Resonance Problems}",
      journal = {Celestial Mechanics},
         year = 1984,
        month = feb,
       volume = {32},
       number = {2},
        pages = {127-136},
          doi = {10.1007/BF01231120},
       adsurl = {https://ui.adsabs.harvard.edu/abs/1984CeMec..32..127B}
}

@ARTICLE{Lee:2002,
       author = {{Lee}, Man Hoi and {Peale}, S.~J.},
        title = "{Dynamics and Origin of the 2:1 Orbital Resonances of the GJ 876 Planets}",
      journal = {\apj},
         year = 2002,
        month = mar,
       volume = {567},
       number = {1},
        pages = {596-609},
          doi = {10.1086/338504},
       adsurl = {https://ui.adsabs.harvard.edu/abs/2002ApJ...567..596L}
}

@ARTICLE{Yang:2024,
       author = {{Yang}, Huan and {Li}, Ya-Ping},
        title = "{Mean-motion resonances with interfering density waves}",
      journal = {\mnras},
         year = 2024,
        month = oct,
       volume = {534},
       number = {1},
        pages = {485-501},
          doi = {10.1093/mnras/stae2097},
archivePrefix = {arXiv},
       eprint = {2309.15694},
 primaryClass = {astro-ph.EP},
       adsurl = {https://ui.adsabs.harvard.edu/abs/2024MNRAS.534..485Y}
}

@ARTICLE{Goldreich:2014,
       author = {{Goldreich}, Peter and {Schlichting}, Hilke E.},
        title = "{Overstable Librations can Account for the Paucity of Mean Motion Resonances among Exoplanet Pairs}",
      journal = {\aj},
         year = 2014,
        month = feb,
       volume = {147},
       number = {2},
          eid = {32},
        pages = {32},
          doi = {10.1088/0004-6256/147/2/32},
archivePrefix = {arXiv},
       eprint = {1308.4688},
 primaryClass = {astro-ph.EP},
       adsurl = {https://ui.adsabs.harvard.edu/abs/2014AJ....147...32G}
}

@ARTICLE{chen:2025,
       author = {{Chen}, Yi-Xian and {Wu}, Yinhao and {Li}, Ya-Ping and {Lin}, Douglas N.~C. and {Alexander}, Richard and {Nayakshin}, Sergei and {Dai}, Fei},
        title = "{Capture and escape of planetary mean-motion resonances in turbulent discs}",
      journal = {\mnras},
         year = 2025,
        month = jun,
       volume = {540},
       number = {2},
        pages = {1998-2007},
          doi = {10.1093/mnras/staf867},
archivePrefix = {arXiv},
       eprint = {2505.13952},
 primaryClass = {astro-ph.EP},
       adsurl = {https://ui.adsabs.harvard.edu/abs/2025MNRAS.540.1998C}
}

@ARTICLE{Emsenhuber:2021,
       author = {{Emsenhuber}, Alexandre and {Mordasini}, Christoph and {Burn}, Remo and {Alibert}, Yann and {Benz}, Willy and {Asphaug}, Erik},
        title = "{The New Generation Planetary Population Synthesis (NGPPS). I. Bern global model of planet formation and evolution, model tests, and emerging planetary systems}",
      journal = {\aap},
         year = 2021,
        month = dec,
       volume = {656},
          eid = {A69},
        pages = {A69},
          doi = {10.1051/0004-6361/202038553},
archivePrefix = {arXiv},
       eprint = {2007.05561},
 primaryClass = {astro-ph.EP},
       adsurl = {https://ui.adsabs.harvard.edu/abs/2021A&A...656A..69E}
}

@ARTICLE{Izidoro:2017,
       author = {{Izidoro}, Andre and {Ogihara}, Masahiro and {Raymond}, Sean N. and {Morbidelli}, Alessandro and {Pierens}, Arnaud and {Bitsch}, Bertram and {Cossou}, Christophe and {Hersant}, Franck},
        title = "{Breaking the chains: hot super-Earth systems from migration and disruption of compact resonant chains}",
      journal = {\mnras},
         year = 2017,
        month = sep,
       volume = {470},
       number = {2},
        pages = {1750-1770},
          doi = {10.1093/mnras/stx1232},
archivePrefix = {arXiv},
       eprint = {1703.03634},
 primaryClass = {astro-ph.EP},
       adsurl = {https://ui.adsabs.harvard.edu/abs/2017MNRAS.470.1750I}
}

@ARTICLE{Fabrycky:2014,
       author = {{Fabrycky}, Daniel C. and {Lissauer}, Jack J. and {Ragozzine}, Darin and {Rowe}, Jason F. and {Steffen}, Jason H. and {Agol}, Eric and {Barclay}, Thomas and {Batalha}, Natalie and {Borucki}, William and {Ciardi}, David R. and {Ford}, Eric B. and {Gautier}, Thomas N. and {Geary}, John C. and {Holman}, Matthew J. and {Jenkins}, Jon M. and {Li}, Jie and {Morehead}, Robert C. and {Morris}, Robert L. and {Shporer}, Avi and {Smith}, Jeffrey C. and {Still}, Martin and {Van Cleve}, Jeffrey},
        title = "{Architecture of Kepler's Multi-transiting Systems. II. New Investigations with Twice as Many Candidates}",
      journal = {\apj},
         year = 2014,
        month = aug,
       volume = {790},
       number = {2},
          eid = {146},
        pages = {146},
          doi = {10.1088/0004-637X/790/2/146},
archivePrefix = {arXiv},
       eprint = {1202.6328},
 primaryClass = {astro-ph.EP},
       adsurl = {https://ui.adsabs.harvard.edu/abs/2014ApJ...790..146F}
}

@ARTICLE{Huang:2023,
       author = {{Huang}, Shuo and {Ormel}, Chris W.},
        title = "{When, where, and how many planets end up in first-order resonances?}",
      journal = {\mnras},
         year = 2023,
        month = jun,
       volume = {522},
       number = {1},
        pages = {828-846},
          doi = {10.1093/mnras/stad1032},
archivePrefix = {arXiv},
       eprint = {2302.03070},
 primaryClass = {astro-ph.EP},
       adsurl = {https://ui.adsabs.harvard.edu/abs/2023MNRAS.522..828H}
}

@ARTICLE{Goldberg:2023,
       author = {{Goldberg}, Max and {Batygin}, Konstantin},
        title = "{Dynamics and Origins of the Near-resonant Kepler Planets}",
      journal = {\apj},
         year = 2023,
        month = may,
       volume = {948},
       number = {1},
          eid = {12},
        pages = {12},
          doi = {10.3847/1538-4357/acc9ae},
archivePrefix = {arXiv},
       eprint = {2211.16725},
 primaryClass = {astro-ph.EP},
       adsurl = {https://ui.adsabs.harvard.edu/abs/2023ApJ...948...12G}
}

@ARTICLE{Hu:2025,
       author = {{Hu}, Zhecheng and {Dai}, Fei and {Zhu}, Wei and {Wang}, Mu-Tian and {Goldberg}, Max and {Lammers}, Caleb and {Masuda}, Kento},
        title = "{Unexpected Near-Resonant and Metastable States of Young Multiplanet Systems}",
      journal = {\apj},
         year = 2025,
        month = dec,
       volume = {995},
       number = {2},
          eid = {206},
        pages = {206},
          doi = {10.3847/1538-4357/ae173c},
archivePrefix = {arXiv},
       eprint = {2510.20185},
 primaryClass = {astro-ph.EP},
       adsurl = {https://ui.adsabs.harvard.edu/abs/2025ApJ...995..206H}
}

@ARTICLE{Mackereth:2019,
       author = {{Mackereth}, J. Ted and {Bovy}, Jo and {Leung}, Henry W. and {Schiavon}, Ricardo P. and {Trick}, Wilma H. and {Chaplin}, William J. and {Cunha}, Katia and {Feuillet}, Diane K. and {Majewski}, Steven R. and {Martig}, Marie and {Miglio}, Andrea and {Nidever}, David and {Pinsonneault}, Marc H. and {Aguirre}, Victor Silva and {Sobeck}, Jennifer and {Tayar}, Jamie and {Zasowski}, Gail},
        title = "{Dynamical heating across the Milky Way disc using APOGEE and Gaia}",
      journal = {\mnras},
         year = 2019,
        month = oct,
       volume = {489},
       number = {1},
        pages = {176-195},
          doi = {10.1093/mnras/stz1521},
archivePrefix = {arXiv},
       eprint = {1901.04502},
 primaryClass = {astro-ph.GA},
       adsurl = {https://ui.adsabs.harvard.edu/abs/2019MNRAS.489..176M}
}

@ARTICLE{Tu:2025,
       author = {{Tu}, Pei-Wei and {Xie}, Ji-Wei and {Chen}, Di-Chang and {Zhou}, Ji-Lin},
        title = "{Age dependence of the occurrence and architecture of ultra-short-period planet systems}",
      journal = {Nature Astronomy},
         year = 2025,
        month = jul,
       volume = {9},
        pages = {995-1006},
          doi = {10.1038/s41550-025-02539-1},
archivePrefix = {arXiv},
       eprint = {2504.20986},
 primaryClass = {astro-ph.EP},
       adsurl = {https://ui.adsabs.harvard.edu/abs/2025NatAs...9..995T}
}

@ARTICLE{Campante:2015,
       author = {{Campante}, T.~L. and {Barclay}, T. and {Swift}, J.~J. and {Huber}, D. and {Adibekyan}, V. Zh. and {Cochran}, W. and {Burke}, C.~J. and {Isaacson}, H. and {Quintana}, E.~V. and {Davies}, G.~R. and {Silva Aguirre}, V. and {Ragozzine}, D. and {Riddle}, R. and {Baranec}, C. and {Basu}, S. and {Chaplin}, W.~J. and {Christensen-Dalsgaard}, J. and {Metcalfe}, T.~S. and {Bedding}, T.~R. and {Handberg}, R. and {Stello}, D. and {Brewer}, J.~M. and {Hekker}, S. and {Karoff}, C. and {Kolbl}, R. and {Law}, N.~M. and {Lundkvist}, M. and {Miglio}, A. and {Rowe}, J.~F. and {Santos}, N.~C. and {Van Laerhoven}, C. and {Arentoft}, T. and {Elsworth}, Y.~P. and {Fischer}, D.~A. and {Kawaler}, S.~D. and {Kjeldsen}, H. and {Lund}, M.~N. and {Marcy}, G.~W. and {Sousa}, S.~G. and {Sozzetti}, A. and {White}, T.~R.},
        title = "{An Ancient Extrasolar System with Five Sub-Earth-size Planets}",
      journal = {\apj},
         year = 2015,
        month = feb,
       volume = {799},
       number = {2},
          eid = {170},
        pages = {170},
          doi = {10.1088/0004-637X/799/2/170},
archivePrefix = {arXiv},
       eprint = {1501.06227},
 primaryClass = {astro-ph.EP},
       adsurl = {https://ui.adsabs.harvard.edu/abs/2015ApJ...799..170C}
}

@ARTICLE{gaia_dr3,
       author = {{Gaia Collaboration} and {Vallenari}, A. and {Brown}, A.~G.~A. and {Prusti}, T. and {de Bruijne}, J.~H.~J. and {Arenou}, F. and {Babusiaux}, C. and {Biermann}, M. and {Creevey}, O.~L. and {Ducourant}, C. and {Evans}, D.~W. and {Eyer}, L. and {Guerra}, R. and {Hutton}, A. and {Jordi}, C. and {Klioner}, S.~A. and {Lammers}, U.~L. and {Lindegren}, L. and {Luri}, X. and {Mignard}, F. and {Panem}, C. and {Pourbaix}, D. and {Randich}, S. and {Sartoretti}, P. and {Soubiran}, C. and {Tanga}, P. and {Walton}, N.~A. and {Bailer-Jones}, C.~A.~L. and {Bastian}, U. and {Drimmel}, R. and {Jansen}, F. and {Katz}, D. and {Lattanzi}, M.~G. and {van Leeuwen}, F. and {Bakker}, J. and {Cacciari}, C. and {Casta{\~n}eda}, J. and {De Angeli}, F. and {Fabricius}, C. and {Fouesneau}, M. and {Fr{\'e}mat}, Y. and {Galluccio}, L. and {Guerrier}, A. and {Heiter}, U. and {Masana}, E. and {Messineo}, R. and {Mowlavi}, N. and {Nicolas}, C. and {Nienartowicz}, K. and {Pailler}, F. and {Panuzzo}, P. and {Riclet}, F. and {Roux}, W. and {Seabroke}, G.~M. and {Sordo}, R. and {Th{\'e}venin}, F. and {Gracia-Abril}, G. and {Portell}, J. and {Teyssier}, D. and {Altmann}, M. and {Andrae}, R. and {Audard}, M. and {Bellas-Velidis}, I. and {Benson}, K. and {Berthier}, J. and {Blomme}, R. and {Burgess}, P.~W. and {Busonero}, D. and {Busso}, G. and {C{\'a}novas}, H. and {Carry}, B. and {Cellino}, A. and {Cheek}, N. and {Clementini}, G. and {Damerdji}, Y. and {Davidson}, M. and {de Teodoro}, P. and {Nu{\~n}ez Campos}, M. and {Delchambre}, L. and {Dell'Oro}, A. and {Esquej}, P. and {Fern{\'a}ndez-Hern{\'a}ndez}, J. and {Fraile}, E. and {Garabato}, D. and {Garc{\'\i}a-Lario}, P. and {Gosset}, E. and {Haigron}, R. and {Halbwachs}, J.-L. and {Hambly}, N.~C. and {Harrison}, D.~L. and {Hern{\'a}ndez}, J. and {Hestroffer}, D. and {Hodgkin}, S.~T. and {Holl}, B. and {Jan{\ss}en}, K. and {Jevardat de Fombelle}, G. and {Jordan}, S. and {Krone-Martins}, A. and {Lanzafame}, A.~C. and {L{\"o}ffler}, W. and {Marchal}, O. and {Marrese}, P.~M. and {Moitinho}, A. and {Muinonen}, K. and {Osborne}, P. and {Pancino}, E. and {Pauwels}, T. and {Recio-Blanco}, A. and {Reyl{\'e}}, C. and {Riello}, M. and {Rimoldini}, L. and {Roegiers}, T. and {Rybizki}, J. and {Sarro}, L.~M. and {Siopis}, C. and {Smith}, M. and {Sozzetti}, A. and {Utrilla}, E. and {van Leeuwen}, M. and {Abbas}, U. and {{\'A}brah{\'a}m}, P. and {Abreu Aramburu}, A. and {Aerts}, C. and {Aguado}, J.~J. and {Ajaj}, M. and {Aldea-Montero}, F. and {Altavilla}, G. and {{\'A}lvarez}, M.~A. and {Alves}, J. and {Anders}, F. and {Anderson}, R.~I. and {Anglada Varela}, E. and {Antoja}, T. and {Baines}, D. and {Baker}, S.~G. and {Balaguer-N{\'u}{\~n}ez}, L. and {Balbinot}, E. and {Balog}, Z. and {Barache}, C. and {Barbato}, D. and {Barros}, M. and {Barstow}, M.~A. and {Bartolom{\'e}}, S. and {Bassilana}, J.-L. and {Bauchet}, N. and {Becciani}, U. and {Bellazzini}, M. and {Berihuete}, A. and {Bernet}, M. and {Bertone}, S. and {Bianchi}, L. and {Binnenfeld}, A. and {Blanco-Cuaresma}, S. and {Blazere}, A. and {Boch}, T. and {Bombrun}, A. and {Bossini}, D. and {Bouquillon}, S. and {Bragaglia}, A. and {Bramante}, L. and {Breedt}, E. and {Bressan}, A. and {Brouillet}, N. and {Brugaletta}, E. and {Bucciarelli}, B. and {Burlacu}, A. and {Butkevich}, A.~G. and {Buzzi}, R. and {Caffau}, E. and {Cancelliere}, R. and {Cantat-Gaudin}, T. and {Carballo}, R. and {Carlucci}, T. and {Carnerero}, M.~I. and {Carrasco}, J.~M. and {Casamiquela}, L. and {Castellani}, M. and {Castro-Ginard}, A. and {Chaoul}, L. and {Charlot}, P. and {Chemin}, L. and {Chiaramida}, V. and {Chiavassa}, A. and {Chornay}, N. and {Comoretto}, G. and {Contursi}, G. and {Cooper}, W.~J. and {Cornez}, T. and {Cowell}, S. and {Crifo}, F. and {Cropper}, M. and {Crosta}, M. and {Crowley}, C. and {Dafonte}, C. and {Dapergolas}, A. and {David}, M. and {David}, P. and {de Laverny}, P. and {De Luise}, F. and {De March}, R.},
        title = "{Gaia Data Release 3. Summary of the content and survey properties}",
      journal = {\aap},
         year = 2023,
        month = jun,
       volume = {674},
          eid = {A1},
        pages = {A1},
          doi = {10.1051/0004-6361/202243940},
archivePrefix = {arXiv},
       eprint = {2208.00211},
 primaryClass = {astro-ph.GA},
       adsurl = {https://ui.adsabs.harvard.edu/abs/2023A&A...674A...1G}
}

\end{CJK*}
\end{document}